%% file: cav2021.tex
\RequirePackage{amsmath}  
\documentclass[runningheads]{llncs}

\usepackage{cabin}  

\usepackage{amssymb}
\usepackage{listings}
\usepackage[table]{xcolor}
\usepackage{hyperref}
\usepackage[normalem]{ulem}
\usepackage{mathtools}
\usepackage{graphicx}
\usepackage{comment}
\usepackage{xspace}
\usepackage{multirow}
\usepackage[rightcaption]{sidecap}
\usepackage[normalem]{ulem} 
\usepackage{enumitem}
\usepackage{relsize}
\usepackage[firstpage]{draftwatermark}
\usepackage[misc]{ifsym}

\usepackage{tikz}
\usetikzlibrary{arrows, positioning, fit, plotmarks}
\usepackage{tikzsymbols}
\usepackage{pgfplots}
\pgfplotsset{compat=1.16}

\input{macros}

\SetWatermarkText{\hspace*{5in}\raisebox{5.in}{\includegraphics[scale=0.8]{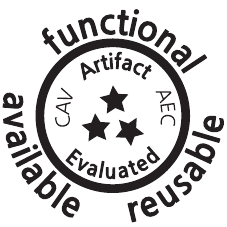}}}

\SetWatermarkAngle{0}

\title{Theory Exploration Powered By Deductive Synthesis}

\authorrunning{Eytan Singher and Shachar Itzhaky}

\author{Eytan Singher\,\Letter \and
    Shachar Itzhaky}
\institute{Technion, Haifa, Israel\\ \email{\{eytan.s,shachari\}@cs.technion.ac.il}}

\begin{document}

\maketitle

\input{00-abstract}

\input{01-intro}
\input{02-overview}
\input{03-prelim}
\input{04-technical}
\input{05-results}
\input{07-related}
\input{09-conclusions}

\clearpage

\bibliography{bib}

\end{document}

%% file: macros.tex
\newif\iffinal
\newif\ifcomments
\commentstrue
\finaltrue

\ifcomments
\newcommand\ES[1]{{\color{orange}[ES: #1]}}
\newcommand\SI[1]{{\color{purple}[SI: #1]}}
\else
\iffinal\else
\newcommand\ES[1]{{}}
\newcommand\SI[1]{{}}
\fi
\fi

\newcommand\Eg{\textit{E.g.\@}\xspace}
\newcommand\ie{\textit{i.e.\@}\xspace}

\newcommand\vs{\textit{vs.\@}\xspace}

\newenvironment{myparagraph}[1]
    {\smallskip\noindent\textbf{#1}~~}{}

\newsavebox{\incog}
\sbox{\incog}{
  \includegraphics[height=5mm,trim=10 10 10 10,clip]
       {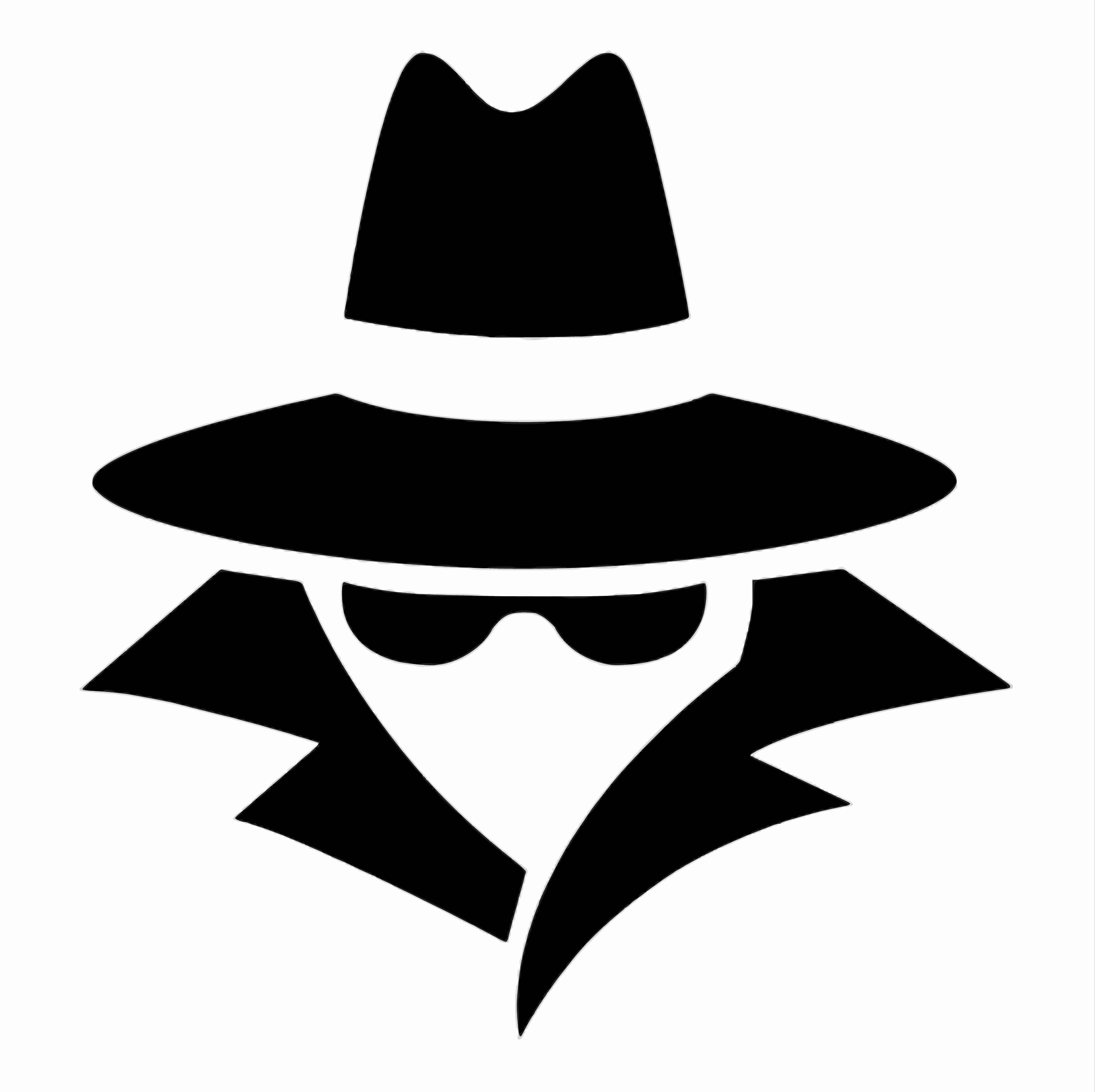}
}

\newcommand\TheSy{{TheSy}\xspace}

\newcommand\operatorstyle[1]{\textrm{\fontfamily{ptm}\selectfont#1}}
\DeclareMathOperator{\cons}{\operatorstyle{::}}

\DeclareMathOperator{\concat}{\operatorstyle{++}}

\newcommand{\lnil}{\mbox{\relscale{0.9}$[\,]$}}

\DeclareMathOperator{\maybeEq}{\overset{?}{\,=\,}}
\DeclareMathOperator{\maybeEqTiny}{\overset{\scriptscriptstyle ?}{\,=\,}}
\DeclareMathOperator{\eqdef}{~\widehat{=}~}

\newcommand\Lang{\mathcal{L}}
\newcommand\Vocab{\mathcal{V}}
\newcommand\Constr{\mathcal{C}}
\newcommand\Eqs{\mathcal{E}}
\newcommand\Rewrite{\mathcal{R}}

\DeclareMathOperator{\rwto}{\overset{.\,}{\rightarrow}}
\newcommand\Terms{\mathcal{T}}

\newcommand\Types{\mathcal{T_Y}}

\newcommand\RewriteRelation[2]{#1 \xrightarrow[]{\Rewrite} #2}
\newcommand\SRewriteRelation[2]{#1 \xleftrightarrow[]{\Rewrite\,} #2}
\newcommand\SRewriteRelationi[2]{#1 \xleftrightarrow[]{\Rewrite_i} #2}
\newcommand\RewriteSys{\{\Rewrite_i\}}

\newcommand\RewriteSysRelation[2]{#1 \xleftrightarrow[]{\RewriteSys} #2}

\newcommand\TRewriteSysRelationSymbol{\xleftrightarrow[]{\RewriteSys}^*}
\newcommand\TRewriteSysRelation[2]{{#1} \,\TRewriteSysRelationSymbol\! {#2}}
\newcommand\TRewriteSysRelationTiny[2]{{#1} \xleftrightarrow[]{\scriptscriptstyle\RewriteSys^*\!\!} {#2}}
\newcommand\TLRewriteSysRelation[2]{#1 {\xleftrightarrow[]{\RewriteSys}^*}\big|_\Terms #2}

\newcommand\TRewriteSysRelationBounded[3]{#1 \underset{\Terms}{\xleftrightarrow[]{\RewriteSys}}^{\!(#3)} #2}
\newcommand\FV{\mathrm{FV}}

\newcommand\pholder[2]{\overset{\scriptscriptstyle #2}{\circ_{\mbox{\relscale{.6}$#1$}}}}  
\newcommand\Symexes[1]{\mathcal{S}^{#1}}

\newcommand\prvbySinline{\vdash^{^{\hspace{-.4em}S\hspace{-.2em}}}}

\newcommand\tfilter{\textrm{filter}}

\newcommand\tdrop{\textrm{drop}}
\newcommand\ttake{\textrm{take}}

\newcommand\tprim[1]{{\textsf{\itshape\relscale{.9}#1}}}
\newcommand\ttrue{\tprim{true}}
\newcommand\tfalse{\tprim{false}}

\newcommand\trev{\textrm{rev}}  

\newcommand\ttype[1]{{\textsf{\relscale{.9}#1}}}
\newcommand\tlist{\ttype{list}}
\newcommand\tbool{\ttype{bool}}

\newcommand\tkeyword[1]{\textrm{\color{blue!70!black}#1}}

\newcommand\tif{\tkeyword{if}}
\newcommand\tthen{\tkeyword{then}}
\newcommand\telse{\tkeyword{else}}
\newcommand\tmatch{\tkeyword{match}}
\newcommand\twith{\tkeyword{with}}

\newcommand\clrterm{\color{blue!65!green}}
\newcommand\clratom{\color{blue!60!green}}
\newcommand\clrsymex{\color{orange!40!red}}
\newcommand\clrhi{\color{red!70!black}}

\newcommand\cmptheories{\textsf{\color{violet!50!white}\smaller{\hspace{1pt}\%\hspace{1pt}}}}

%% file: 00-abstract.tex
\begin{abstract}
This paper presents a symbolic method for automatic theorem generation based on deductive inference. 
Many software verification and reasoning tasks require proving complex logical properties; 
coping with this complexity is generally done by declaring and proving relevant sub-properties.
This gives rise to the challenge of discovering useful sub-properties that can assist the automated proof process.
This is known as the \emph{theory exploration} problem, and so far, predominant solutions that emerged rely on evaluation using concrete values.
This limits the applicability of these theory exploration techniques to complex programs and~properties.

In this work, we introduce a new symbolic technique for theory exploration, capable of (offline) generation of a library of lemmas from a base set of inductive data types and recursive definitions.
Our approach introduces a new method for using abstraction to overcome the above limitations, combining it with deductive synthesis to reason about abstract values.
Our implementation has shown to find more lemmas than prior art, avoiding redundant lemmas (in terms of provability), while being faster in most cases.
This new abstraction-based theory exploration method is a step toward applying theory exploration to software verification and synthesis.

\keywords{Theory Exploration  \and Synthesis
    \and Automatic Theorem Proving.}
\end{abstract}

%% file: 01-intro.tex
\section{Introduction}
\label{intro}

Most forms of software verification and synthesis rely on some form of logical reasoning to complete their task.
Whether it is checking pre- and post-conditions, deriving specifications for sub-problems~\cite{PLDI2015:Feser,POPL2016:Albarghouthi}, or equivalence reduction~\cite{VMCAI2019:Smith}, these methods rely on assumptions from both the input and relevant background knowledge.
Domain-specific knowledge can reinforce these methods, whether via the design of a domain-specific language~\cite{PLDI2001:Xiong,HPEC2000:Moura,CACM2018:Ragan-Kelley}, specialized decision procedures~\cite{TODAES2012:Milder}, or decomposing specifications~\cite{ACM2015:Polozov}.
While hand-crafted techniques can treat whole classes of programs, every library or module contributes a collection of new primitives, requiring tweaking or extending these methods.
Automatic formation of background knowledge can enable effortless treatment of such libraries and programs.

In the context of verification tools, such as Dafny~\cite{LPAIR2010:Rustan} and Leon~\cite{SCALA13:Blanc}, as well as interactive proof assistants, such as 
Coq~\cite{Coq:manual} and Isabelle/HOL~\cite{Book2002:Nipkow},
background knowledge is typically given as a set of \emph{lemmas}.
Usually, these libraries of lemmas (\ie the background knowledge) are created by human engineers and researchers who are tasked with formulating them and proving their correctness.
When a proof or verification task requires auxiliary lemmas missing from the existing background knowledge, 
the user is required to add and prove it, sometimes repeating this process until the proof is trivial or can be found automatically.
For example, both Dafny and Leon fail to prove that addition is associative and commutative from first principles---based on an algebraic construction of the natural numbers.
However, when given knowledge of these properties (\ie encoded as lemmas: $(x + y) + z = x + (y + z)$ and $x + y = y + x$)%
\footnote{In fact, these properties are hard-wired into decision procedures for linear integer arithmetic in SMT solvers.}, they readily prove composite facts such as
$(x + 5) + y = 5 + (x + y)$.

A possible solution is to eagerly generate valid lemmas, and to do so automatically, offline,
as a precursor to any work that would be built on top of the library.
This paradigm is known as \emph{theory exploration}~\cite{Journal2002:Buchberger,JAL2006:Buchberger}, and differs from the common conjecture generation approach (in theorem provers and SMT solvers~\cite{VMCAI2015:Reynolds}) that is guided by a proof goal.
As opposed to using proof goal as the basis for discovering sub-goals, when eagerly generating lemmas there is a vast space of possible lemmas to consider.
Currently, two main approaches exist for filtering candidate conjectures, counterexample-based and observational equivalence-based~\cite{JAR2010:Johansson,2018AISC:Einarsdottir,CICM2014:Johansson,ECEASST2015:Valbuena}.
These filtering techniques are all based on testing and therefore require automatic creation of concrete examples.

Testing with concrete values allows for fast evaluation and filtering of terms when the data types involved are simple.
However, when scaling to larger data types and function types it becomes a bottleneck of the theory exploration process.
Previous research effort has revealed that testing-based discovery is sensitive to the number and size of type definitions occurring in the code base.
For example, QuickSpec, which is based on QuickCheck (as are all the existing testing-based theory exploration methods), employs a heuristic to restrict the set of types allowed in terms in order to make the checker's job easier.
Compound data types such as lists can be nested up to two levels (lists of lists, but not lists of lists of lists).
This presents an obstacle towards scaling the approach to real software libraries, since ``\textit{QuickCheck's size control
interacts badly with deeply nested types} [...] \textit{will generate extremely large test data.}''~\cite{JFP2017:Smallbone}

Following are two example scenarios that attempt to represent cases from software systems where structured data types and complicated APIs exist:
(i)~A series of tree data-types $T_i$ where each $T_i$ is a tree of height i with i children of type $T_{i-1}$, and the base case is an empty tree.
Creating concrete examples for $T_i$ will be resource expensive, as each tree has $O(i!)$ nodes, and each node requires a value.
(ii)~An ADT (Algebraic Data Type) $A$ with multiple fields where each can contain a large amount of text or other ADTs, and a function over $A$ that only accesses one of the fields.
Even if evaluating the function is fast, fully creating $A$ is expensive and will impact the theory exploration run-time.

This paper presents a new symbolic theory exploration approach that takes advantage of the characteristics of induction-based proofs.
To overcome the blowup in the space of possible values, we make use of \emph{symbolic values}, which contain interpreted symbols, uninterpreted symbols, or a mixture of the two. 
Conceptually, each symbolic value is an abstraction representing (infinitely) many possible values. 
This means that preexisting knowledge on the symbolic value can be applied without fully creating interpreted values.
Still, when necessary, uninterpreted values can be expanded, creating larger symbolic values, thus refining the abstraction, and facilitating the necessary computation.
We focus on the formation of \emph{equational} theories, that is, lemmas that curtail the equivalence of two terms, with universal quantification over all free variables.

We show that our symbolic method for theory exploration is more applicable and faster in many different scenarios than state-of-the-art.
As an example, given standard definitions for the list functions: $\clrterm\concat~\tdrop~\ttake~\tfilter$
our method proves facts that were not found by current state-of-the-art such as:

\centerline{$
\renewcommand\arraystretch{1.3}
\begin{array}{c}
{(\ttake~i~\mathit{xs})}~\concat~{(\tdrop~i~\mathit{xs})} = \mathit{xs} \\
\tfilter~p~(\mathit{xs}\,\concat\,\mathit{ys}) = (\tfilter~p~\mathit{xs})~\concat~(\tfilter~p~\mathit{ys})
\end{array}
$}

\medskip
\begin{myparagraph}{Main contributions}
This paper provides the following contributions:
\begin{itemize}[topsep=2pt, leftmargin=1.5em]
    \item A system for \emph{theory synthesis} using symbolic values to take advantage of value abstraction.
    Our implementation, \TheSy, can discover more lemmas than were found by testing-based tools, while being faster in most cases.
    \item A technique to compare universally quantified terms using term rewriting techniques and a given set of lemmas, called \emph{symbolic observational equivalence} (SOE). 
    SOE overapproximates term equalities deducible by the given lemmas (\ie, will find more equalities), thus can be used for equality reduction in context of uninterpreted values, enabling fully symbolic reasoning over a large set of terms.
    \item An evaluation of our theory exploration system on a set of benchmarks for induction proofs taken from CVC4~\cite{VMCAI2015:Reynolds} and TIP 2015~\cite{CICM2015:Claessen}, specifically the IsaPlanner benchmarks~\cite{ITP2010:Johansson}.
    We compare our implementation with a current leading theory exploration system, Hipster~\cite{2018AISC:Einarsdottir}, using a novel metric. This metric is insensitive to the amount of found lemmas, but rather measures their usefulness in the context of theorem proving.
\end{itemize}
\end{myparagraph}

%% file: 02-overview.tex
\section{Overview}
\label{overview}

Our theory exploration method, named \TheSy (Theory Synthesizer, pronounced \emph{Tessy}), is based on syntax-guided enumerative synthesis.
Similarly to previous approaches \cite{JFP2017:Smallbone,ICAD2013:Claessen,ITP2017:Johansson}, TheSy generates a comprehensive set of
terms from the given vocabulary and looks for pairs that seem equivalent.
Notably, \TheSy employs deductive reasoning based on term rewriting systems to propose these pairs
by extrapolating from a set of known equalities, employing a relatively lightweight (but unsound) reasoning procedure.
The proposed pairs are passed as equality conjectures to a theorem prover capable of reasoning by induction.

\begin{figure}[t]
\resizebox{\textwidth}{!}{
\begin{tikzpicture}[>=latex, stage/.style={inner xsep=5pt, inner ysep=9pt}, coupled/.style={draw=black!40!white, dotted, thick}, node distance=7mm]
    \node(term-gen)[stage, draw, align=center]
        { Term \\ Generation \\ (SyGuE) };
    \node(infer)[stage, draw, align=center, right=of term-gen]
        { Conjecture \\ Inference \\ (SOE) };
    \node(screen)[stage, draw, align=center, right=of infer]
        { Conjecture \\ Screening \\ ~ };
    \node(induction)[stage, draw, align=center, right=of screen]
        { Induction \\ Prover \\ {\small(cong. closure)}};
        
    \node(infer+screen)[coupled, fit=(infer) (screen)] {};
        
    \draw[->] (term-gen) -- (infer);
    \draw[->] (infer) -- (screen);
    \draw[->] (screen) -- (induction);
    \draw[->] (induction.east) -- ++(.4,0) -- ++(0,1.25)
              -| node[above,pos=.25] {\small iterative deepening} (term-gen);
    \draw[->] (induction.south) -- ++(0,-.25)
             -| node[below, pos=.25] {\small augment knowledge} 
             coordinate[pos=.25](midpoint) (term-gen);
    \draw[->] (midpoint -| infer) -- (infer);
    \draw[->] (midpoint -| screen) -- (screen);
    \draw[<-] (term-gen.west) -- ++(-5mm,0)
        node[anchor=east, align=center, inner sep=0, xshift=3mm, yshift=-2mm] { \small base \\ \small knowledge };
    \draw[->] (induction.east) -- ++(7mm,0)
        node[anchor=west, align=center, inner sep=1pt, xshift=-3mm, yshift=-2mm] { \small new \\ \small knowledge };
\end{tikzpicture}
}
\caption{\TheSy system overview: breakdown into phases, with feedback loop.}
\label{overview:sys-flow}
\end{figure}
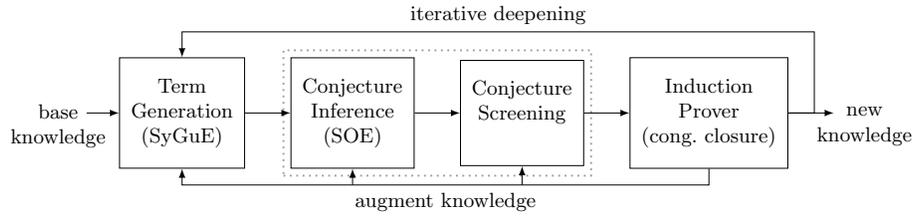

The process (as shown in \autoref{overview:sys-flow}) is separated into four stages.
These stages work in an iterative deepening fashion and are dependent on the results of each other. A short description is given to help the reader understand their context later on.
\begin{enumerate}[leftmargin=1.2em]
    \item \textbf{Term Generation.~} Build symbolic terms of increasing depth, based on the given vocabulary.
    Use known equalities for pruning via equivalence reduction.
    \item \textbf{Conjecture Inference.~} Evaluate terms on symbolic inputs, and apply deductive inference to extract new equalities, thus forming conjectures.
    \item \textbf{Conjecture Screening.~} Some of the conjectures, even valid ones, are special cases of known equalities or are trivially implied by them. We deem these conjectures redundant.
    TheSy culls such conjectures before continuing to prove the rest.
    \item \textbf{Induction Prover.~} The prover attempts to prove conjectures that passed screening using a normal induction scheme derived from algebraic data structure definitions in the given vocabulary.
    Conjectures that were successfully proven are then declared \emph{lemmas} and added to the known equalities.
\end{enumerate}

The phases are run iteratively in a loop, where each iteration deepens the generated terms and, hence, the discovered lemmas.
These lemmas are fed back to earlier phases;
this form of feedback contributes to discovering more lemmas thanks to several factors:
\begin{enumerate}[label=(\roman*), align=left, labelsep=0pt,labelwidth=2em]
\item Conjecture inference is dependent upon known equalities. Additional equalities enable finding new conjectures.
\item Accurate screening by merging equivalence classes based on known equalities. 
\item The prover is based on known equalities with a congruence closure procedure. 
The more lemmas are known to the system, the more lemmas become provable by this method.
\item Term generation benefits from the equivalence reduction, avoiding duplicate work for equivalent terms. 
\end{enumerate}

\begin{figure}[t]
\[
\begin{array}{rll}
  \Vocab & = & \{ \begin{array}[t]{l@{\quad}l}
         \lnil       & \tlist~T, \\
         {\cons}     & T \to \tlist~T \to \tlist~T, \\
         {\concat}   & \tlist~T \to \tlist~T \to \tlist~T,  \\
         \tfilter & (T\to\tbool)\to\tlist~T\to\tlist~T~\}
         \end{array}
          
  \hspace{3em}
  \Constr ~=~ \{~\lnil, {\cons}~\} \\
         \\
  \Eqs    & = & \{\begin{array}[t]{l@{\quad}ll}
          \lnil \concat l = l, &
          (x\cons xs)\concat l = x\cons(xs\concat l), \\
          \tfilter~p~\lnil = \lnil,& 
          \tfilter~p~(x\cons xs) = \tif~p\,x~
          	\tthen~x\cons\tfilter~p~xs~
            \telse~\tfilter~p~xs &\}
          \end{array}
\end{array}
\]
\caption{An example input to \TheSy.}
\label{overview:input-state}
\end{figure}

\begin{paragraph}{Running example.}
To illustrate \TheSy's theory exploration procedure,
we introduce a simple running example based on a list ADT. The input given to \TheSy is shown in \autoref{overview:input-state};
it consists of a vocabulary $\Vocab$ (of which $\Constr$ is a subset of ADT constructors) and a set of known equalities $\Eqs$.
The vocabulary $\Vocab$ contains the canonical list constructors $\clratom\lnil$ and $\clratom\cons$, and two basic list operations $\clratom\concat$ (concatenate) and $\clratom\tfilter$.
The equalities $\Eqs$ consist of the definitions of the latter two.
\end{paragraph}

\medskip
At a very high level, the following process is about to take place:
\TheSy generates symbolic terms representing length-bound lists, \textit{e.g.}, $\clrsymex\lnil$, $\clrsymex[v_1]$, $\clrsymex[v_2,v_1]$.
Then, it will evaluate all combinations of function applications, up to a small depth, using these symbolic terms as arguments.
If these evaluations yield common values for all possible assignments, the two application terms yielding them are conjectured to be equal.
Since the evaluated expressions contain symbolic values, their result is a symbolic value. 
Comparing such symbolic values is done via congruence closure-based reasoning; we call this process \emph{symbolic observational equivalence}, by way of analogy to observational equivalence~\cite{CAV2103:Albarghouthi} that is carried out using concrete values.

Out of the conjectures computed using symbolic observational equivalence, \TheSy selects minimal ones according to a combined metric of compactness and generality.
These are passed to a prover that employs both congruence closure and induction to verify the correction of the lemmas for \emph{all} possible list values.

Some lemmas that \TheSy can discover this way are:
\[
\renewcommand\arraystretch{1.2}
\begin{array}{c}
\tfilter~p~(\tfilter~p~l) = \tfilter~p~l
 \hspace{1.5cm}
l_1 \,{\concat}\, (l_2 \,{\concat}~ l_3) = (l_1 \,{\concat}~ l_2) \,{\concat}\, l_3 \\
\tfilter~p~l_1 \concat \,\tfilter~p~l_2 =
\tfilter~p~(l_1{\concat}\,l_2)
\end{array}
\]

As briefly mentioned, our system design relies on congruence closure-based reasoning over universally quantified first-order formulas with uninterpreted functions.
Congruence closure is weak but fast and constitutes one of the core procedures in SMT solvers~\cite{JACM1980:Nelson,IC2007:Nieuwenhuis}.
On top of that, universally-quantified assumptions~\cite{TACAS2017:Barbosa} are handled by formulating them as rewrite rules and applying some depth-bounded term rewriting as described in \autoref{prelim:term-rewriting}.
Additionally, \TheSy implements a simple case splitting mechanism that enables reasoning on conditional expressions.
Notably, this procedure \emph{cannot} reason about recursive definitions since such reasoning routinely requires the use of induction.
To that end, \TheSy is geared towards discovering lemmas that can be proven by induction; a lemma is considered useful if it cannot be proven from existing lemmas by congruence closure alone, that is, without induction.
Discovering such lemmas and adding them to the background knowledge evidently increases the reasoning power of the prover, since at least the fact of their own validity becomes provable, which it was not before.

%% file: 03-prelim.tex
\section{Preliminaries}
This work relies heavily on term rewriting techniques, which is employed across multiple phases of the exploration.
Term rewriting is implemented efficiently using equality graphs (e-graphs).
In this section, we present some minimal background of both, which will be relevant for the exploration procedure described later.

\subsection{Term Rewriting Systems}
\label{prelim:term-rewriting}

Consider a formal language $\Lang$ of terms over some vocabulary of symbols.
We use the notation $\Rewrite = t_1 \rwto t_2$ to denote a rewrite rule from $t_1$ to $t_2$.
For a (universally quantified) semantic equality law $t_1 = t_2$, we would normally create \emph{both}
$t_1 \rwto t_2$ and $t_2 \rwto t_1$.
We refrain from assigning a direction to equalities since we do not wish to restrict the procedure
to strongly normalizing systems, as is traditionally done in frameworks based on the Knuth-Bendix
algorithm~\cite{AR1983:Knuth}.
Instead, we define equivalence when a sequence of rewrites can identify the terms in either direction.
A small caveat involves situations where $\FV(t_1)\neq\FV(t_2)$, that is, one side of the equality
contains variables that do not occur on the other.
We choose to admit only rules $t_i \rwto t_j$ where $\FV(t_i)\supseteq\FV(t_j)$, because when
$\FV(t_i)\subset \FV(t_j)$, applying the rewrite would have to create new symbols for the unassigned
variables in $t_j$, which results in a large growth in the number of symbols and typically makes
rewrites much slower as a result.

\medskip
This slight asymmetry is what motivates the following definitions.

\begin{definition}\label{screening:rw-equiv}
Given a rewrite rule $\Rewrite = t_1 \rwto t_2$, we define a corresponding relation
$\RewriteRelation{}{}$ such that $\RewriteRelation{s_1}{s_2} \iff s_1=C[t_1\sigma] \land s_2=C[t_2\sigma]$
for some context $C$ and substitution $\sigma$ for the free variables of $t_1, t_2$.
(A \emph{context} is a term with a single hole, and $C[t]$ denotes the term obtained by filling the
hole with $t$.)
\end{definition}

\begin{definition}\label{screening:symmetric}
Given a relation $\RewriteRelation{}{}$ we define its symmetric closure:

\centering
$\SRewriteRelation{t_1}{t_2} \iff \RewriteRelation{t_1}{t_2} ~\lor~ \RewriteRelation{t_2}{t_1}$
\end{definition}

\begin{definition}\label{screening:rw-set-equiv}
Given a set of rewrite rules $G_\Rewrite=\RewriteSys$, we define a relation as union of the relations of the rewrites: $\RewriteSysRelation{}{} \eqdef \bigcup_{i}\SRewriteRelationi{}{}$.\\
In the sequel, we will mostly use its reflexive transitive closure, $\TRewriteSysRelationSymbol$.
\end{definition}

The relation $\TRewriteSysRelationSymbol$ is reflexive, transitive, and symmetric,
so it is an equivalence relation over $\Lang$.
Under the assumption that all rewrite rules in $\RewriteSys$ are semantics preserving, for any equivalence class $[t] \in \Lang \big/ {\TRewriteSysRelationSymbol}\!$, all terms belonging to $[t]$ are definitionally equal.
However, since $\Lang$ may be infinite, it is essentially impossible to compute $\TRewriteSysRelationSymbol\!$.
Any algorithm can only explore a finite subset $\Terms \subseteq \Lang$, and in turn,
construct a subset of $\TRewriteSysRelationSymbol\!$.

\subsection{Compact Representation Using Equality Graphs}
\label{screening:representation}

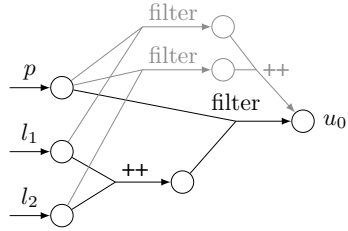
\begin{SCfigure}[1.18][t]
\scalebox{.9}{
\input{img/peg}
}
\caption{An e-graph representing the expression
$\clrterm\textrm{filter}~p~(l_1 \concat l_2)$ (dark)
and the equivalent expression
$\clrterm\textrm{filter}~p~l_1 \concat \textrm{filter}~p~l_2$ (light).}
\label{prelim:peg}
\end{SCfigure}

In order to be able to cover a large set of terms $\Terms$, we need a
compact data structure that can efficiently represent many terms.
Normally, terms are represented by their ASTs
(Abstract Syntax Trees), but as there would
be many instances of common subterms among the terms of $\Terms$, 
this would be highly inefficient.
Instead, we adopt the concept of equality graphs (e-graphs) from automated theorem proving~\cite{JACM2005:Detlefs},
which also saw uses in compiler optimizations and program synthesis~\cite{POPL2009:Tate,PLDI2015:Panchekha,PLDI2020:Nandi}, in which context they are known as Program Expression Graphs (PEGs).
An e-graph is essentially a hypergraph where each vertex represents a set
of equivalent terms (programs), and labeled, directed hyperedges
represent function applications.
Hyperedges therefore have exactly one target and zero or more sources,
which form an ordered multiset (a vector, basically).
Just to illustrate, the expression
$\clrterm\tfilter~p~(l_1 \concat l_2)$
will be represented by the nodes and edges shown in dark in \autoref{prelim:peg}.
The nullary edges represent the constant symbols ($p$, $l_1$, $l_2$), and the node $u_0$ represents the entire term.
The expression 
$\clrterm\textrm{filter}~p~l_1 \concat \textrm{filter}~p~l_2$, which is equivalent, is represented by the light nodes and edges, and the equivalence is captured by sharing of the node $u_0$.

When used in combination with a rewrite system $\RewriteSys$, each
rewrite rule is represented as a premise pattern $P$ and a conclusion pattern $C$.
Applying a rewrite rule is then reduced to searching the e-graph for the search pattern and obtaining a substitution $\sigma$ for the free variables of $P$.
The result term is then obtained by substituting the free variables of $C$ using $\sigma$.
This term is added to the same equivalence class as the matched term (\ie{} $P\sigma$), meaning they will both have the same root node.
Consequently, a single node can represent a set of terms
exponentially large in the number of edges, all of which will always be equivalent modulo 
$\TRewriteSysRelationSymbol$.

In addition, since hyperedges always represent functions, a situation may arise
in which two vertices represent the same term:
This happens if two edges $\bar{u}\xrightarrow{\scriptscriptstyle f}v_1$ and $\bar{u}\xrightarrow{\scriptscriptstyle f}v_2$
are introduced by $\RewriteSys$ for $v_1\neq v_2$.
In a purely functional setting, this means that $v_1$ and $v_2$ are equal.
Therefore, when such duplication is found,
it is beneficial to \emph{merge} $v_1$ and $v_2$, eliminating the duplicate hyperedge.
The e-graph data structure therefore supports a vertex merge operation and a congruence closure-based
transformation~\cite{POPL2021:Willsey}
that finds vertices eligible for merge to keep the overall graph size small.
This procedure can be quite expensive, so it is only run periodically.

%% file: img/peg.tex
\begin{tikzpicture}[>=latex,cp/.style={draw,circle},
        nuc/.style={inner sep=0, outer sep=0},
        ly/.style={black!45!white}]
    \node(u1)[cp] {};
    \node(u2)[cp, below=6mm of u1] {};
    \node(u3)[cp, below=6mm of u2] {};
    \node(p)[left=6mm of u1] {};
    \node(l1)[left=6mm of u2] {};
    \node(l2)[left=6mm of u3] {};
    \draw[->] (p) -- node[above] {$p$} (u1);
    \draw[->] (l1) -- node[above] {$l_1$} (u2);
    \draw[->] (l2) -- node[above] {$l_2$} (u3);
    \node(e1)[nuc, right=6mm of u2, yshift=-4.5mm] {};
    \node(u4)[cp, right=8mm of e1] {};
    \draw[->] (u2) -- (e1.center) -- (u4);
    \draw (u3) -- (e1.center);
    \node[above right=0 of e1, inner sep=2pt] {$\concat$};
    \node(e2)[nuc, right=6mm of u4, yshift=9mm] {};
    \node(u5)[cp, right=8mm of e2] {};
    \draw[->] (u1) -- (e2.center) -- (u5);
    \draw (u4) -- (e2.center);
    \node[above=0 of e2, inner sep=4pt] {filter};
    \node[right=0 of u5] {$u_0$};
    
    \node(e3)[nuc, right=of u1, yshift=9mm] {};
    \node(u6)[cp, ly, right=of e3] {};
    \draw[ly,->] (u1) -- (e3.center) -- (u6);
    \draw[ly] (u2) -- (e3.center);
    \node[above right=0 of e3, inner sep=2pt, ly] {filter};
    \node(e4)[nuc, below=6mm of e3] {};
    \node(u7)[cp, ly, right=of e4] {};
    \draw[ly,->] (u1) -- (e4.center) -- (u7);
    \draw[ly] (u3) -- (e4.center);
    \node[above right=0 of e4, inner sep=2pt, ly] {filter};
    \node(e5)[nuc, right=3mm of u7] {};
    \draw[ly,->] (u6) -- (e5.center) -- (u5);
    \draw[ly] (u7) -- (e5.center);
    \node[right=0 of e5, inner sep=2pt, ly] {$\concat$};
    
\end{tikzpicture}

%% file: 04-technical.tex
\section{Theory Synthesis}
\label{technical}

In this section, we go into a more detailed description of the phases of theory synthesis and explain how they are combined within an iterative deepening loop.
To simplify the presentation, we describe all the phases first, then explain how
the output from the last phase is fed back to the next iteration to complete a
feedback loop.
We continue with the input from the running example in \autoref{overview} (\autoref{overview:input-state}) and dive deeper by showing intermediate states encountered during the execution of \TheSy on this input.
Throughout the execution, \TheSy maintains a state, consisting of the following elements:
\begin{itemize}
  \item $\Vocab$, a sorted vocabulary 
  \item $\Constr\subseteq\Vocab$, a subset of constructors for some or all of the types
  \item $\Eqs$, a set of equations initially consisting only of the
    definitions of the (non-constructor) functions in $\Vocab$
  \item $\Terms$, a set of terms, initially containing just atomic terms corresponding to symbols from $\Vocab$.
\end{itemize}

\subsection{Term Generation}
\label{overview:term-generation}

The first step is to generate a set of terms over the vocabulary $\Vocab$.
For the purpose of generating universally-quantified conjectures, we introduce a set of uninterpreted symbols, which we will call \emph{placeholders}.
Let $\Types$ be the set of types occurring as the type of some argument of a function symbol in $\Vocab$.
For each type $\tau$ occurring in $\Vocab$ we generate placeholders $\pholder{i}{\tau~}$, two for each type (we will explain later why two are enough).
These placeholders, together with all the symbols in $\Vocab$, constitute the terms at depth $0$.

At every iteration of deepening, \TheSy uses the set of terms generated so far, and the (non-nullary) symbols of $\Vocab$, to form new terms by placing existing ones in argument positions.
For example, with the definitions from \autoref{overview:input-state},
we will have terms such as these at depths 1 and 2:
\bgroup
\renewcommand\arraystretch{1.4}
\arrayrulecolor{black!25}
\begin{equation}\label{overview:example-terms}
\begin{array}{l@{\qquad}c@{\qquad}c@{\qquad}}
  {\scriptstyle 1} &
  \tfilter\,\pholder{1}{\!T\!\to\tbool}~\pholder{1}{\tlist\,T} &
  \pholder{1}{\tlist\,T} \concat \pholder{2}{\tlist\,T}
\\ \hline
  {\scriptstyle 2} &
  \lnil ~\concat~
  \tfilter\,\pholder{1}{\!T\!\to\tbool}~\pholder{1}{\tlist\,T} &
  \pholder{1}{\tlist\,T} \concat 
    ~(\tfilter\,\pholder{1}{\!T\!\to\tbool}~\pholder{2}{\tlist\,T})
\\
  & 
  \tfilter\,\pholder{1}{\!T\!\to\tbool}~
    (\pholder{1}{\tlist\,T} \concat \pholder{2}{\tlist\,T})  &
  (\tfilter\,\pholder{1}{\!T\!\to\tbool}~\pholder{1}{\tlist\,T})
    ~\concat ~
    (\tfilter\,\pholder{1}{\!T\!\to\tbool}~\pholder{2}{\tlist\,T})
\end{array}
\end{equation}
\egroup

It is easy to see that 
$\clrterm\tfilter\pholder{1}{\!T\!\to\tbool}\pholder{1}{\tlist\,T}$ and $\clrterm\lnil \concat
  \tfilter\pholder{1}{\!T\!\to\tbool}\pholder{1}{\tlist\,T}$
are equivalent in any context; this follows directly from the definition of $\concat$, available as part of $\Eqs$. 
It is therefore acceptable to discard one of them without affecting completeness.
\TheSy does not discard terms---since they are merged in the e-graph, there is no need to---rather, it chooses the smaller term as representative when it needs one.
This sort of \emph{equivalence reduction} is present, in some way or another, in many automated reasoning and synthesis tools.

To formalize the procedure of generating and comparing the terms,
in an attempt to discover new equality conjectures,
we introduce the concept of \emph{Syntax Guided Enumeration} (SyGuE).
SyGuE is similar to Syntax Guided Synthesis (SyGuS for short~\cite{DSSE2015:Alur}) in that they both use a formal definition of a language to find program terms solving a problem.
They differ in the problem definition: while SyGuS is defined as a search for a correct program over the well-formed programs in the language,
SyGuE is the sub-problem of iterating over \emph{all distinct} programs in the language.
SyGuS solvers may be improved using a smart search algorithm, while SyGuE solvers need an efficient way to eliminate duplicate terms, which may depend on the definition of program equivalence.
We implement our variant of SyGuE, over the equivalence relation $\TRewriteSysRelationTiny{\!}{\!\!}$, using the aforementioned e-graph:
by applying and re-applying rewrite rules,
provably equivalent terms are naturally \emph{merged} into hyper-vertices, representing equivalence classes.

\subsection{Conjecture Inference \& Screening}
\label{overview:conjecture}

Of course, in order to discover \emph{new} conjectures,
we cannot rely solely on term rewriting based on $\Eqs$.
To find more equivalent terms, \TheSy carries on to generate a second set of terms, called \emph{symbolic examples},
this time using only the constructors $\Constr \subset \Vocab$
and uninterpreted symbols for leaves.
This set is denoted $\Symexes{\tau}$, where $\tau$ is an algebraic datatype participating in $\Vocab$
(if several such datatypes are present, one $\Symexes{\tau}$ per type is constructed).
The depth of the symbolic examples (i.e. depth of applied constructors) is also bounded, but it is independent of the current term depth and does not increase during execution.
For example, using the constructors of $\tlist~T$ with an example depth of $2$,
we obtain the symbolic examples 
$\Symexes{\tlist\,T\!} = \{
  {\clrsymex\lnil}, {\clrsymex v_1{\cons}\lnil}, 
  {\clrsymex v_2{\cons} v_1{\cons}\lnil}\}$, 
corresponding to lists of length up to $2$ having arbitrary element values.
Intuitively, if two terms are equivalent for all possible assignments of symbolic examples to $\pholder{i}{\tlist\,T}$,
then we are going \emph{hypothesize} that they are equivalent for all
list values.
This process is very similar to observational equivalence as used by program
synthesis tools~\cite{CAV2103:Albarghouthi,Notices2013:Udupa}, but since it uses the symbolic value terms instead of concrete
values, we dub it \emph{symbolic observational equivalence} (SOE).


\medskip
Consider, for example, the simple terms $\clrterm\pholder{1}{\tlist\,T}$
and $\clrterm\pholder{1}{\tlist\,T}\!\concat\lnil$.
In placeholder form, none of the rewrite rules derived from $\Eqs$ applies, so it cannot be determined that these terms are, in fact, equivalent.
However, with the symbolic list examples above, the following rewrites are enabled:
\[
\TRewriteSysRelation
{\clrterm{\clrsymex\lnil}\concat\lnil}{\clrsymex\lnil}   \hspace{1.5em} 
\TRewriteSysRelation
{\clrterm{\clrsymex v_1{\cons}{\lnil}}\concat\lnil}
  {\clrsymex v_1{\cons}{\lnil}}
 \hspace{1.5em}
\TRewriteSysRelation
{\clrterm{\clrsymex v_2{\cons} v_1{\cons}\lnil}\concat\lnil}
  {\clrsymex v_2{\cons} v_1{\cons}\lnil}
\]

A similar case can be made for the two bottom terms in \eqref{overview:example-terms}.
For symbolic values $l_1,l_2\in\Symexes{\tlist\,T\!}$,
it can be shown that
\[
\TRewriteSysRelation
  {\clrterm\tfilter\pholder{1}{T\!\to\tbool}({\clrsymex l_1} 
    \concat\, {\clrsymex l_2})\,}
  {\clrterm{(\tfilter\pholder{1}{T\!\to\tbool}{\clrsymex l_1})} 
    \concat {(\tfilter\pholder{1}{T\!\to\tbool}{\clrsymex l_2})}}
\]

In fact, it is sufficient to substitute for $\pholder{1}{\tlist\,T}$, while \emph{leaving} $\pholder{2}{\tlist\,T}$ \emph{alone, uninterpreted}:
e.g., 
$
\TRewriteSysRelation
  {\clrterm\tfilter\pholder{1}{T\!\to\tbool}({\clrsymex\lnil}
    \concat \pholder{2}{\tlist\,T})}
  {\clrterm{(\tfilter\pholder{1}{T\!\to\tbool}{\clrsymex\lnil})}
    \concat 
    {(\tfilter\pholder{1}{T\!\to\tbool}\pholder{2}{\tlist\,T})}}
$.
This reduces the number of equivalence checks significantly,
and is more than a mere heuristic:
since we are going to rely on a prover that proceeds by applying induction to one of the arguments, it makes perfect sense to only bound that argument.
If computation is blocked on the second argument, we would prefer to first infer an auxiliary lemma first, then use it to discover the blocked lemma later.
See \autoref{overview:lemma-seeding} below for an idea of when this situation arises.

The attentive reader may notice that the cases of
$\clrsymex v_1{\cons}\lnil$ and
$\clrsymex v_2{\cons}v_1{\cons}\lnil$
are a bit more involved:
to proceed with the rewrite of $\clrterm\tfilter$, the expressions
$\clrterm\pholder{1}{T\!\to\tbool}\,v_{\scriptscriptstyle 1}$,
$\clrterm\pholder{1}{T\!\to\tbool}\,v_{\scriptscriptstyle 2}$
must be resolved to either $\ttrue$ or $\tfalse$.
However, the predicate $\pholder{1}{T\!\to\tbool}$ as well as the arguments $v_{\scriptscriptstyle 1,2}$ are uninterpreted.
In this case, \TheSy is required to perform a \emph{case split} in order to enable the rewrites and unify the symbolic terms separately in each of the resulting four ({\small $2^2$}) cases.
Notice that leaving $\pholder{1}{T\!\to\tbool}$ uninterpreted means that the cases are only split when evaluation is blocked by one or more rewrite rule applications, potentially saving some branching.
The following steps are then carried out for each case.

\TheSy applies all the available rewrite rules to the entire e-graph, containing all the terms and symbolic examples.
For every two terms $t_1,t_2$ such that for all viable substitutions $\sigma$ of placeholders to symbolic examples of the corresponding types, $t_1\sigma$ and $t_2\sigma$ were shown equal---that is, ended up in the same equivalence class of the e-graph---the conjecture $t_1\maybeEqTiny t_2$ is emitted. \Eg, in the case of the running example:

\[
{\clrterm
  \tfilter\pholder{1}{T\!\to\tbool}(
    \pholder{1}{\tlist\,T} \concat 
    \pholder{2}{\tlist\,T})}
~\maybeEq~
{\clrterm
  {(\tfilter\pholder{1}{T\!\to\tbool}
    \pholder{1}{\tlist\,T})}
    \,\concat\,
  {(\tfilter\pholder{1}{T\!\to\tbool}
    \pholder{2}{\tlist\,T})}}
\]

In the presence of multiple cases, the results are intersected, so that a conjecture is emitted only if it follows from all the cases.

\begin{paragraph}{\bf Screening.}
Generating all the pairs according to the above criteria potentially creates many ``obvious'' equalities, which are valid propositions, but do not contribute to the overall knowledge and just clutter the prover's state.
For example,
\[
{\clrterm
\tfilter\pholder{1}{T\!\to\tbool}
 ({\pholder{1}{\tlist\,T}} \concat\, {\pholder{2}{\tlist\,T}})} ~\maybeEq~
{\clrterm
\tfilter\pholder{1}{T\!\to\tbool}
  \big(\,{\pholder{1}{\tlist\,T}} \concat\,
  {(\lnil \concat {\pholder{2}{\tlist\,T}})}\big)}
\]
which follows from the definition of ${\concat}$ and has nothing to do with
$\clrterm\tfilter$. 
The synthesizer avoids generating such candidates, by
choosing at most one term from every equivalence class
of placeholder-form terms induced during the term generation phase.
If both sides of the equality conjecture belong to the same equivalence class, the conjecture is dropped altogether.
\end{paragraph}

The conjectures that remain are those equalities $t_1\maybeEqTiny t_2$ where $t_1$ and $t_2$ got merged for all the assignments $\Symexes{\tau}$ to some $\pholder{1}{\tau~\,}$, and, furthermore, $t_1$ and $t_2$ themselves \emph{were not} merged in placeholder form,
prior to substitution.
Such conjectures, if true, are guaranteed to increase the knowledge represented by $\Eqs$ as (at least) the equality $t_1=t_2$ was not previously provable using term rewriting and congruence closure.

\subsection{Induction Prover}
\label{overview:induction}

For practical reasons, the prover employs the following induction tactic:
\begin{itemize}
  \item \emph{Structural} induction based on the provided 
    constructors ($\Constr$).
  \item The \emph{first} placeholder of the inductive type is selected as the decreasing argument.
   \item Exactly \emph{one} level of induction is attempted for each candidate.
\end{itemize}

The reasoning behind this design choice is that for every multi-variable
term, \emph{e.g.} $\clrterm\pholder{1}{\tlist\,T} \concat \pholder{2}{\tlist\,T}$,
the synthesizer also generates the symmetric counterpart
$\clrterm\pholder{2}{\tlist\,T} \concat \pholder{1}{\tlist\,T}$.
So electing to perform induction on $\pholder{1}{\tlist\,T}$ does not impede
generality.

In addition, if more than one level of induction is needed, the proof can
(almost) always be revised by factoring out the inner induction as an auxiliary lemma.
Since the synthesizer produces \emph{all} candidate equalities, that inner
lemma will also be discovered and proved with one level of induction.
Lemmas so proven are added to $\Eqs$ and are available to the prover, so that
multiple passes over the candidates can gradually grow the set of provable equalities.

When starting a proof, the prover never needs to look at the base case,
as this case has already been checked during conjecture inference.
Recall that placeholders $\pholder{1}{\tau~}$ are instantiated with bounded-depth expressions using the constructors of $\tau$,
and these include all base cases (non-recursive constructors) by default.
For the example discussed above, the case of
${\clrterm\tfilter\pholder{1}{T\!\to\tbool}({\clrsymex\lnil}
    \concat \pholder{2}{\tlist\,T})}
=    
  {\clrterm{(\tfilter\pholder{1}{T\!\to\tbool}  
  {\clrsymex\lnil})}
    \,\concat\,
    {(\tfilter\pholder{1}{T\!\to\tbool}\pholder{2}{\tlist\,T})}}$
has been discharged early on, otherwise the conjecture would not have come to pass.
The prover then turns to the induction step, which is pretty routine but
is included in \autoref{overview:induction-example} for completeness of the presentation.
\begin{figure}[t]
\renewcommand\arraystretch{1.3}
\newcommand\ih{~~~~(\textit{IH}\,)}
\[
\begin{array}{l@{\quad}l}
\textit{Assume} & 
\tfilter~p~(\mathit{xs} \concat l_1) =
 \tfilter~p~\mathit{xs} \concat \tfilter~p~l_1
 \\
\textit{Prove} &
 \tfilter~p~((x \cons \mathit{xs}) \concat l_1) =
 \tfilter~p~(x \cons \mathit{xs}) \concat \tfilter~p~l_1
\\
\textit{via} & (1)~~ \tfilter~p~((x \cons \mathrm{xs}) \concat l_1) 
                 = \tfilter~p~(x \cons (\mathrm{xs} \concat l_1)) \\
             & (2)~~ \qquad = \tmatch~(p\,x)~\twith~
                \begin{array}[t]{@{}l@{}}
                  \ttrue\Rightarrow 
                       x\cons\tfilter~p~(\textit{xs} \concat l_1) \\
                  \tfalse\Rightarrow 
                              \tfilter~p~(\textit{xs} \concat l_1)
                \end{array} \\
    \ih      & (3)~~ \qquad = \tmatch~(p\,x)~\twith~
                \begin{array}[t]{@{}l@{}}
                  \ttrue\Rightarrow 
                       x\cons {(\tfilter~p~\mathit{xs} \concat 
                                \tfilter~p~l_1)} \\
                  \tfalse\Rightarrow 
                              \tfilter~p~\mathit{xs} \concat 
                                \tfilter~p~l_1 
                \end{array} \\
             & (4)~~\tfilter~p~(x \cons \mathit{xs}) \concat 
                    \tfilter~p~l_1 \\
             & \hspace{1.3cm} = \big(\tmatch~(p\,x)~\twith~
                \begin{array}[t]{@{}l@{}}
                  \ttrue\Rightarrow 
                       x\cons \tfilter~p~\mathit{xs} \\
                  \tfalse\Rightarrow  \tfilter~p~\mathit{xs}\big)
                      \concat \tfilter~p~l_1
                \end{array} \\
             & (5)~~ \qquad = \tmatch~(p\,x)~\twith~
                \begin{array}[t]{@{}l@{}}
                  \ttrue\Rightarrow 
                       x\cons {(\tfilter~p~\mathit{xs} \concat 
                                \tfilter~p~l_1)} \\
                  \tfalse\Rightarrow 
                              \tfilter~p~\mathit{xs} \concat 
                                \tfilter~p~l_1 
                \end{array} \\[-1.5em]
\square &
\end{array}
\]
\vspace{-5mm}
\caption{Example proof by induction based on congruence closure and
  case splitting.}
  \label{overview:induction-example}
\end{figure}

It is worth noting that the conjecture inference, screening and induction phases
utilize a common reasoning core based on rewriting and congruence closure.
In situations where the definitions include conditions such as
$\clrterm\tmatch~p\,x$ in \autoref{overview:induction-example} (in this case, desugared from $\clrterm\tif~p\,x$), the prover also performs automatic case split
and distributes equalities over the branches.
Details and specific optimizations are described in \autoref{evaluation}.

\begin{paragraph}{Speculative generalization.}
When the prover receives a conjecture with multiple occurrences of a placeholder, \textit{e.g.}
${\clrterm {\pholder{1}{\tlist\,T}} \concat 
   {(\pholder{2}{\tlist\,T} \concat \pholder{1}{\tlist\,T})}}
~\maybeEq~
 {\clrterm {({{\pholder{1}{\tlist\,T}} \concat 
   {\pholder{2}{\tlist\,T}})} \concat \pholder{1}{\tlist\,T}}}$,
it is designed to first speculate a more general form for it by replacing the multiple occurrences with fresh placeholders.
Recall that in \autoref{overview:term-generation} we argued that two placeholders of each type is going to be sufficient;
this is the mechanism that enables it.
There is more than one way to generalize a given conjecture: for this example, there are two ways (up to alpha-renaming):
\[
{\clrterm {\pholder{1}{\tlist\,T}} \concat 
   {(\pholder{2}{\tlist\,T} \concat {\clrhi\pholder{3}{\tlist\,T}})}}
\maybeEq
 {\clrterm {({{\pholder{1}{\tlist\,T}} \concat 
   {\pholder{2}{\tlist\,T}})} \concat {\clrhi\pholder{3}{\tlist\,T}}}}
\hspace{2em}
{\clrterm {\pholder{1}{\tlist\,T}} \concat 
   {(\pholder{2}{\tlist\,T} \concat {\clrhi\pholder{3}{\tlist\,T}})}}
\maybeEq
 {\clrterm {({{\clrhi\pholder{3}{\tlist\,T}} \concat 
   {\pholder{2}{\tlist\,T}})} \concat \pholder{1}{\tlist\,T}}}
\]
The prover must attempt both. Failing that, it would fall back to the original conjecture.
Formally, given an equality conjecture $\clrterm{s = t}$ we can consider an assignment $\sigma$ such that $\clrterm{ r = s\sigma, q = t\sigma}$;
where the original conjecture uses an assignment with only two values per type.
The prover thus must iterate through different assignments $\sigma_i$ with more possible values per type, and attempt to prove a new conjecture $\clrterm{r\sigma_i = q\sigma_i}$.
This incurs more work for the prover but is well worth its cost compared to a-priori generation of terms with three placeholders. 
\end{paragraph}

\subsection{Looping Back}

The equations obtained from \autoref{overview:induction} are fed back in
four different but interrelated ways.
The first, inner feedback loop is from the induction prover to itself:
the system will attempt to prove the smaller lemmas first, so that when
proving the larger ones, these will already be available as part of $\Eqs$.
This enables more proofs to go through.
The second feedback loop uses the lemmas obtained to filter out proofs that are no longer needed.
The third, outer loop is more interesting: as equalities are made into
rewrite rules, additional equations may now pass the inference phase,
since the symbolic evaluation core can equate more terms based on this
additional knowledge.
The fourth resonates with the third, applying the new rewrite rules acts as an equality reduction mechanism, reducing the number of hyperedges added to the e-graph during term generation.

It is worth noting that while concrete observational equivalence uses
a trivially simple equivalence checking mechanism with the trade-off
that it may generate many incorrect equalities,
our \emph{symbolic} observational equivalence is conservative in the
sense that a symbolic value may represent infinitely many concrete
inputs, and only if the synthesizer can \emph{prove} that two terms will
evaluate to equal values on \emph{all} of them, by way of constructing
a small proof, are they marked as equivalent.
This means that some actually-equivalent terms may be ``blocked'' by
the inference phase, which cannot happen when using concrete values---%
but also means that having additional inference rules ($\Eqs$) can improve this equivalence checking, 
potentially leading to more discovered lemmas.
This property of \TheSy is appealing because it allows an explored theory to evolve from basic lemmas to more complex ones.

\begin{example}[Lemma seeding]\label{overview:lemma-seeding}
To understand this last point, consider the standard definition of list reversal for the $\tlist$ datatype:
\[
\begin{array}{rcl}
  \trev~\lnil & = & \lnil \\
  \trev~(x\cons xs) & = & \trev~xs \concat {(x\cons\lnil)}
\end{array}
\]

Given the terms $t_1 = {\clrterm \trev~({\pholder{1}{\tlist\,T}} \concat {\pholder{2}{\tlist\,T}})}$
and $t_2 = {\clrterm {\trev{\pholder{2}{\tlist\,T}}} \,\concat\, {\trev{\pholder{1}{\tlist\,T}}}}$,
symbolic observational equivalence with the assignments
$\{\pholder{1}{\tlist\,T} \mapsto \Symexes{\tlist\,T}\}$
fails to unify them.
This is due to $\concat$ being defined by induction on its first argument,
hence, \textit{e.g.}---
\[
\begin{array}{l@{\quad}c@{\quad}l}
 {\clrterm\trev~({\clrsymex v_2\cons v_1\cons\lnil}
   \concat \pholder{2}{\tlist\,T})} & \to^* &
 {\clrterm\big(\trev\pholder{2}{\tlist\,T} \concat 
   {(v_1\cons\lnil)}\big) \concat {(v_2\cons\lnil)}} \\
 {\clrterm {\trev~{\pholder{2}{\tlist\,T}}} \concat\,
  {\trev~{\clrsymex v_2\cons v_1\cons\lnil}}} & \to^* &
 {\clrterm\trev\pholder{2}{\tlist\,T} \concat 
   {(v_1\cons v_2\cons\lnil)}}
\end{array}
\]

Without the associativity property of $\concat$, it would not be possible to show that these symbolic values are equivalent, so the conjecture
$t_1\maybeEqTiny t_2$ will not even be generated.
Luckily, having proven
${\clrterm {\pholder{1}{\tlist\,T}} \concat 
   {(\pholder{2}{\tlist\,T} \concat \pholder{3}{\tlist\,T})}}
~\maybeEq~
 {\clrterm {({{\pholder{1}{\tlist\,T}} \concat 
   {\pholder{2}{\tlist\,T}})} \concat \pholder{3}{\tlist\,T}}}$,
these rewrites are ``unblocked'', so that the equality can be conjectured and ultimately proven.
\end{example}

One caveat is that whenever $\Eqs$ is updated by the addition of a new lemma, some of the previously emitted conjectures may consequently become redundant.
Moreover, conjectures that were passed to the prover before but failed validation may now succeed, and new ones may be emitted in the generation phase.
To take these into account, the actual loop performed by \TheSy is a bit more involved than has been described so far.
For each term depth, \TheSy performs all phases as described, but each time a lemma is discovered \TheSy re-runs the conjecture generation, screening, and prover phases.
Only when no more conjectures are available does \TheSy increase the term depth and generate new terms.

%% file: 05-results.tex
\section{Evaluation}
\label{evaluation}

We implemented \TheSy in Rust, using the e-graph manipulation library \emph{egg}~\cite{POPL2021:Willsey}.
\TheSy accepts definitions in SMTLIB-2.6 format~\cite{TR2017:Barrett},
based on the UF theory (uninterpreted functions),
limited to universal quantifications.
Type declarations occurring in the input are collected and comprise $\Vocab$;
universal equalities form $\Eqs$ and are translated into rewrite rules (either uni- or bidirectional, as explained in \autoref{prelim:term-rewriting}).
Then SyGuE is performed on $\Vocab$, generating candidate conjectures using SOE.
SyGuE uses \emph{egg} for equivalence reduction, and SOE uses it for comparing symbolic values. Conjectures are then dismissed using \TheSy's induction-based prover.
This is done in an iterative deepening loop.

\begin{paragraph}{Case split\quad}
Both SOE and the prover use a case splitting mechanism; This mechanism detects when rewriting cannot match due to an opaque value (an uninterpreted symbol), and applies case splitting according to the constructors of relevant ADTs.
However, doing so for every rule is too costly and, in most cases, redundant --- \TheSy generates a variety of terms, so if one term is blocked due to an uninterpreted symbol, another one exists with a symbolic example instead.
A situation where this is \emph{not} the case is when \emph{multiple} uninterpreted symbols block the rewrite (recall that \TheSy only substitutes one placeholder per term with symbolic examples). 
To illustrate, consider the case in \autoref{overview:induction-example} where both the list $\clrterm x::xs$ and $\clrterm p\,x$ are used in match expressions, therefore a case split is needed by ${\clrterm p\,x} \in \{\ttrue,\tfalse\}$.
Therefore, \TheSy only performs case splitting for rewrite rules that require multiple match patterns but only one is blocked.

The splitting mechanism itself, operates by copying the e-graph and applying the term rewriting logic separately for each case. 
Each copy then yields a partition of the existing equivalence classes. 
These partitions are intersected between all cases, and each of the resulting intersections lead to merging of equivalence classes in the original e-graph.
It is worth noting that \TheSy never needs to backtrack a case split it has elected to apply.
As a consequence, execution time is not exponential in the total number of case splits performed, only in the nesting level of such splits (which is bounded by 2 in our experiments). 
\end{paragraph}

We compare \TheSy to the most recent and closely related theory exploration system, Hipster~\cite{CICM2014:Johansson}---which is
based on random testing (backed by QuickSpec~\cite{JFP2017:Smallbone}) with proof automations from and frontend in Isabelle/HOL~\cite{Book2002:Nipkow}.
Hipster represents the culmination of several works on existing theory exploration (see~\autoref{related}).
Both systems generate a set of proved lemmas as output, each such set encompassing a conceptual volume of knowledge that was discovered automatically.
We note that the same knowledge can be represented in various ways, so directly comparing the sets of lemmas is going to be meaningless.

\subsection{Evaluating Theory Exploration Quality}

We define a comparison method for two theory exploration systems $A$ and $B$ starting from a common initial theory (defined as a set of closed formulas) $\mathcal{T}$.
As a metric for the quality and efficacy of results obtained from theory exploration, and, therefore, their perceived usefulness, we use the notion of \emph{knowledge} (inspired by ``knowledge base'' in Theorema~\cite{Journal2002:Buchberger}).
A theory $\mathcal{T}$ in a given logical proof system induces a collection of attainable knowledge, $\mathcal{K}_\mathcal{T} = \left\{\varphi ~\rule[-2.5pt]{0.5pt}{9.5pt}~ \mathcal{T}\vdash\varphi\right\}$,
that is, characterized by the set of (true) statements that can be proven based on $\mathcal{T}$.
In practice, a ``pure'' notion of knowledge based on provability is impractical, because most interesting logics are undecidable,
and automated proving techniques cannot feasibly find proofs for all true statements.
We, therefore, parameterize knowledge relative to a \emph{prover} --- a procedure that always terminates and can prove a subset of true statements.
Termination can be achieved by restricting the space of proofs by either size or resource bounds.
We say that $\mathcal{T}
\vdash^{^{\hspace{-.4em}S\hspace{-.2em}}} \varphi$ when a prover, $S$, is able to verify the validity of $\varphi$ in a theory $\mathcal{T}$.
A more realistic characterization of knowledge would then be
$\mathcal{K}_\mathcal{T}^S = \big\{\varphi ~{\rule[-3pt]{0.5pt}{11.5pt}}~
\mathcal{T}\prvbySinline\varphi\big\}$.
Assuming that the prover $S$ is fixed, a theory $\mathcal{T}'$ is said to \emph{increase knowledge} over $\mathcal{T}$ when
$\mathcal{K}_{\mathcal{T}'}^S \supset \mathcal{K}_{\mathcal{T}}^S$.

We utilize the notion of $\mathcal{K}_\mathcal{T}^S$ described above to test the knowledge gained by $A$ against that of $B$, and vice versa.
We take the set of lemmas $\mathcal{T}_A$ generated by $A$ and check whether it is subsumed by $\mathcal{T}_B$, generated by $B$, by checking whether $\mathcal{T}_A\subseteq \mathcal{K}_{\mathcal{T}\cup\mathcal{T}_B}^S$;
we then carry out the same comparison with the roles of $A$ and $B$ reversed.
A working assumption is that both $A$ and $B$ include some mechanism for screening redundant conjectures.
That is, a component that receives the current set of known lemmas $T_i$ and a conjecture $\varphi$ and decides whether the conjecture is redundant.
It is important to choose $S$ such that whenever $A$ (or $B$) discards $\varphi$, due to redundancy, it holds that $\varphi \in \mathcal{K}_{T_i}^S$.

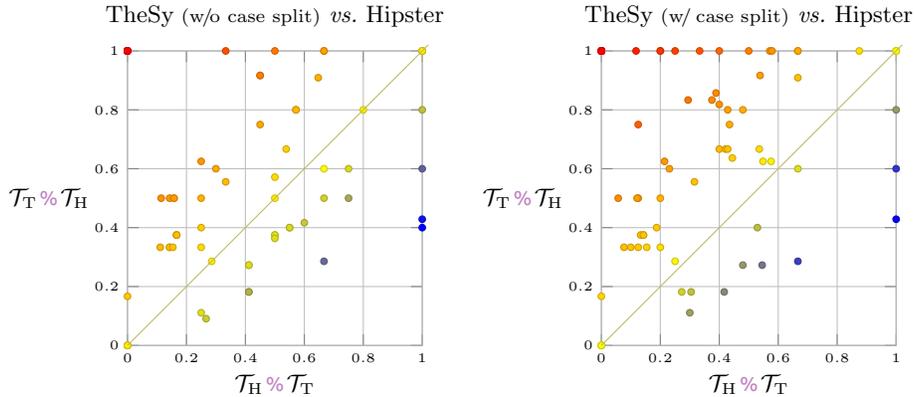
\begin{figure}[t]
    \input{img/scatter-plot}
\vspace{-2em}  
    \caption{A scatter plot showing the ratio of lemmas in theories discovered by each tool that were subsumed by the theory discovered by its counterpart ($\tthesy$ = \TheSy, $\thipster$ = Hipster). Each point represents a single test case.
    The vertical axis shows how many of the lemmas discovered by Hipster were subsumed by those discovered by \TheSy, and the horizontal axis shows the converse.}
    \label{results:ratio_chart}
\end{figure}

Incorporating the solver into the comparison makes the evaluation resistant to large amounts of trivial lemmas, as they will be discarded by A or B.
It is still possible for some lemmas to be ``better'' than others, so knowledge is not uniformly distributed; this is hard to quantify, though.
A few possible measures of usefulness come to mind, such as lemma utilization in a task (such as proof search), proof complexity, or matching to a given context, but given just the exploration task, there is not sufficient information to apply them.
A first approximation is to consider the discovered lemmas themselves, \ie,
$\mathcal{T}_A \cup \mathcal{T}_B$, as representing proof objectives.
In doing so, we pit $A$ and $B$ in direct contest with one another.
We choose this avenue because it is straightforward to apply, admitting that it may be inaccurate in some cases.

To evaluate our approach and its implementation, we run both \TheSy and Hipster
on functional definitions collected from the TIP 2015 benchmark suite~\cite{CICM2015:Claessen}, specifically the IsaPlanner~\cite{ITP2010:Johansson} benchmarks (85 benchmarks in total), for compatibility between the two systems.
TIP benchmarks also contain goal propositions, but for the purpose of evaluating the exploration technique, these are redacted.
This experiment uses the simple rewrite-driven congruence-closure decision procedure with a case split mechanism in the role of the solver, $S$, occurring in the definition of knowledge $\mathcal{K}$.
Hipster uses Isabelle/HOL's simplifier as a conjecture redundancy filtering mechanism, which is in itself a simple rewrite-driven decision procedure, therefore $S$ provides a suitable comparison.
We compute the portion of lemmas found by Hipster that were provable (by $S$) from \TheSy's results and vice versa.
In other words, we check the ratio given by $| \mathcal{T}_A \cap \mathcal{K}_{\mathcal{T}\cup\mathcal{T}_B}^S | \,/\, | \mathcal{T}_A |$, which we denote $\mathcal{T}_B\cmptheories\mathcal{T}_A$, in both directions.
\autoref{results:ratio_chart} displays the ratios, where each point represents a single test case.
Points above the diagonal line represent test cases where \TheSy's ratio was higher and for points under the line Hipster's ratio was higher.
We conduct this experiment twice: Once with the case-splitting mechanism of \TheSy turned off for its exploration, and once with it turned on. (Hipster does not have such a switch as it always generates concrete values.)
The reason for this is that case splitting increases the running time significantly (as we show next), so we want to evaluate its contribution to the discovery of lemmas.
Comparing the two charts, while \TheSy performs reasonably well compared to Hipster without case splitting (in 48 out of the 85 \TheSy's ratio was better and equal in 12), enabling it leads to a clear advantage (in 65 out of the 85 \TheSy's ratio was better and equal in 6).

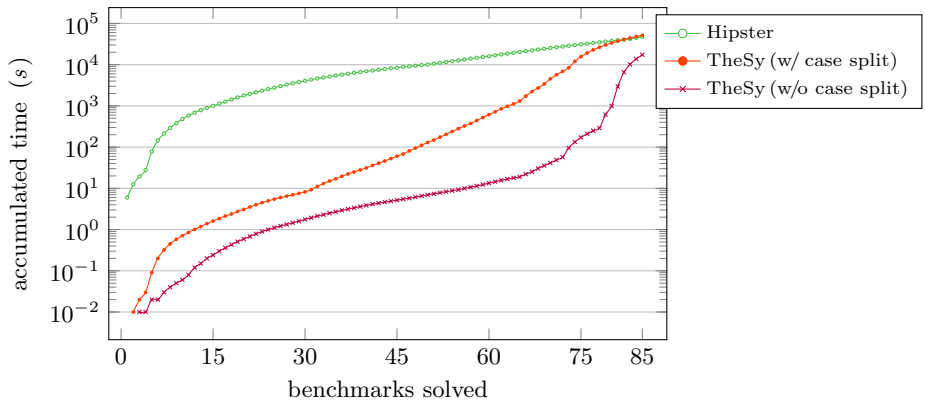
\begin{figure}[t]
\centering
\input{img/cactus-plot.tex}
\vspace{-2em}
\caption{Time to fully explore the 85 IsaPlanner benchmarks.
A full exploration is considered one where either all terms up to the depth bound have been enumerated or a timeout of 1hr has been reached.
The $y$ axis shows the amount of time needed to complete the first $x$ benchmarks, when they are sorted from shortest- to longest-running.
(Time scale is logarithmic; lower is better.)}
\label{results:time-perc-table}
\end{figure}

\subsubsection{Performance}

To compare runtime efficiency, we consider the time it took to fully explore the IsaPlanner test suite.
We consider an exploration ``full'' when it has finished enumerating all the terms, and associated candidate conjectures, up to the depth bound ($k=2$)\footnote{Our experience shows that choosing larger $k$s greatly affects the run-time, but does not lead to many useful lemmas.} with \TheSy or size bound with Hipster ($s=7$), and check them; or when a timeout of one hour is reached, whichever is sooner.
We then sort the benchmarks from shortest- to longest-running for each of the tools, and report the accumulated time to explore the first $i$ benchmarks ($i=1..85$).
The results are shown in the graph in \autoref{results:time-perc-table}, for Hipster, \TheSy with case split disabled, and \TheSy with case split enabled.
In both configurations, \TheSy is very fast for the lower percentiles, but begins to slow down, due to case splitting, towards the end of the line.
To illustrate, in the 25th percentile \TheSy was {\raise.17ex\hbox{$\scriptstyle\sim$}}380 times faster (0.48s \vs 182.47s); in the 50th percentile, {\raise.17ex\hbox{$\scriptstyle\sim$}}57 times faster (5.28s \vs 305.37s); and in the 75th percentile, {\raise.17ex\hbox{$\scriptstyle\sim$}}6 times faster (141.24 to 883.8).
Overall \TheSy took 51.6K seconds and Hipster 47.1K, meaning Hipster was {\raise.17ex\hbox{$\scriptstyle\sim$}}1.1 times faster.
It is evident from the chart that case splitting is largely responsible for the longer execution times. Without case splitting, \TheSy is much faster, and completes all 85 benchmarks in less time than it takes Hipster.
Of course, in that mode of operation, \TheSy finds fewer lemmas (as shown in \autoref{results:ratio_chart}), but is still superior to Hipster.
Future work needs to focus on improving the case-splitting mechanism, similar to their treatment in SAT and SMT, allowing \TheSy to deal with such theories more efficiently.

\subsection{Efficacy to automated proving}

While the mission statement of \TheSy is solely to provide lemmas based on core theories,
we wish to claim that such discovered theories are beneficial toward proving theorems in general, based on the same core theory.
We used a collection of benchmarks for induction proofs used by CVC4~\cite{VMCAI2015:Reynolds}, and conducted the following experiment: First, the proof goals are skipped and only the symbol declarations and provided axioms are used to construct an input to \TheSy.
Then, whenever a new lemma is discovered and passes through the prover, we also attempt to prove the goal---utilizing the same mechanism used for vetting conjectures.
As soon as the latter goes through, the exploration process is aborted, and all lemmas collected are discarded.
The experiments are thus independent across the individual benchmarks.

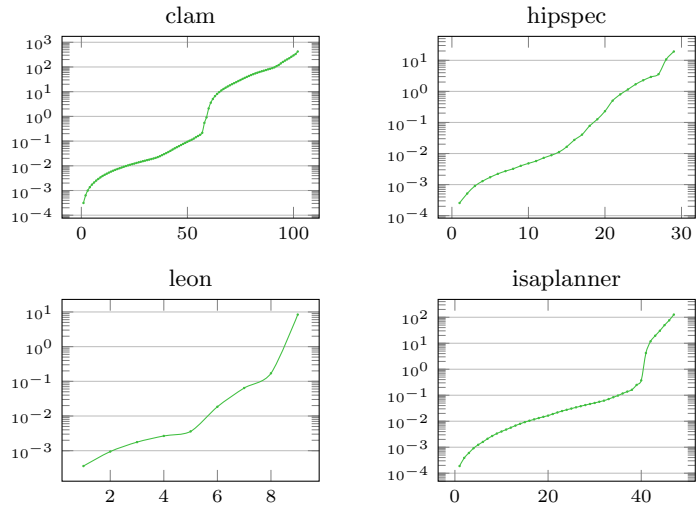
\begin{figure}[t]
    \centering
    \input{img/cvc4-benchmarks/cactus-plot}
    \caption{Accumulated time-to-solve for each of the benchmark suites from the CVC4 collection.
    The $y$ axis shows the amount of time needed to complete the first $x$ (successful) proofs, when benchmarks are sorted from shortest- to longest-running.}
    \label{results:solve-accum}
\end{figure}

Even though this setting is unfavorable to \TheSy---because it does not take advantage of the fact that theory exploration can be done offline, then its results re-used for proofs over the same core theory---we report considerable success in solving these benchmarks.
Out of the 311 benchmarks, our theory exploration + simple-minded induction was able to prove 187 (with a 5-minute timeout, same as in the original CVC4 experiments).
For comparison, Z3 and CVC4 (without conjecture generation) were able to prove 75 and 70 of them, respectively.
This shows that the majority of instances were not solvable without the use of induction.
CVC4 with its conjecture generation enabled was able to solve 260 of them.
\autoref{results:solve-table} shows the number of successful proofs achieved for each of the four suites.
\autoref{results:solve-accum} shows the accumulated time required for the benchmarks; the vast majority of the success cases occur early on, because in some cases a rather small auxiliary lemma is all that is needed to make the proof go through.

\begin{table}[t]
\centering
\input{tables/cvc4-results}
\vspace{5pt}
\caption{Results of the CVC4 benchmark suite (number of successful proofs in each category).}
\label{results:solve-table}
\end{table}

%% file: img/scatter-plot.tex
\newcommand\tthesy{\textrm{T}}
\newcommand\thipster{\textrm{H}}

\newcommand\fwith{w\hspace{-1pt}/\hspace{-1pt}\xspace}
\newcommand\fwithout{w\hspace{-1pt}/\hspace{-1pt}o\xspace}

\begin{tikzpicture}
\coordinate (O) at (0,0);
\begin{axis}[
    title={\TheSy {\smaller~(\fwithout case split)} \vs Hipster},
    at=(O),
    draw={black!30!white},
    width=5.5cm, height=5.5cm,
    grid=both,
    xmin=0, xmax=1,
    ymin=0, ymax=1,
    ticklabel style={font=\tiny},
    x label style={at={(axis description cs:0.5,-0.07)},anchor=north},
    y label style={at={(axis description cs:-0.1,.5)},rotate=-90,anchor=east},
    xlabel={$\mathcal{T}_{\thipster} \cmptheories
             \mathcal{T}_{\tthesy}$},
    ylabel={$\mathcal{T}_{\tthesy} \cmptheories
             \mathcal{T}_{\thipster}$}
]
\addplot[  
    only marks,
    scatter,
    scatter src=y-x,
    point meta min=-.6, point meta max=1,
    mark size=1.25pt]
table[x=t<h,y=h<t]{img/scatter-plot-no-cs.dat.txt};
\end{axis}
\draw[color=olive!60!white] (O) -- +(4cm, 4cm);


\coordinate (O) at (6.3cm,0);
\begin{axis}[
    title={\TheSy {\smaller~(\fwith case split)} \vs Hipster},
    at=(O),
    draw={black!30!white},
    width=5.5cm, height=5.5cm,
    grid=both,
    xmin=0, xmax=1,
    ymin=0, ymax=1,
    ticklabel style={font=\tiny},
    x label style={at={(axis description cs:0.5,-0.07)},anchor=north},
    y label style={at={(axis description cs:-0.1,.5)},rotate=-90,anchor=east},
    xlabel={$\mathcal{T}_{\thipster} \cmptheories
             \mathcal{T}_{\tthesy}$},
    ylabel={$\mathcal{T}_{\tthesy} \cmptheories
             \mathcal{T}_{\thipster}$}
]
\addplot[  
    only marks,
    scatter,
    scatter src=y-x,
    point meta min=-.5, point meta max=1,
    mark size=1.25pt]
table[x=t<h,y=h<t]{img/scatter-plot-with-cs.dat.txt};
\end{axis}
\draw[color=olive!60!white] (O) -- +(4cm, 4cm);
\end{tikzpicture}

%% file: img/cactus-plot.tex
\begin{tikzpicture}
  \begin{axis}[
  		width=9cm,
        height=6cm,
        xmin=-2,
        xmax=89,
        xtick distance=15,
        extra x ticks={85},
        ymode=log,
        ytick distance=10^1,
        ymajorgrids=true,
        xlabel={benchmarks solved},
        ylabel={accumulated time\, 
           (\textit{s}\hspace{.75pt})},
           mark size=.5pt,
        legend style={
          anchor=north west,
          font={\fontsize{7.5pt}{7.5pt}\selectfont}
        },
        legend image post style={mark size=2pt, scale=0.8},
        legend cell align={left}]
        
    \addplot[smooth,mark=*,mark size=.6pt,mark options={fill=white},green!50!gray] plot 
      table[y=hipster] {img/cactus-plot-with-cs.dat.txt};
    \addlegendentry{Hipster}

    \addplot[smooth,mark=*,orange!50!red] plot 
       table[y=thesy] {img/cactus-plot-with-cs.dat.txt};
    \addlegendentry{TheSy\,%
    {\smaller(w\hspace{-1pt}/\hspace{-1pt} case split)}}

    \addplot[smooth,mark=x,mark size=1.1pt,purple] plot 
       table[y=thesy] {img/cactus-plot-no-cs.dat.txt};
    \addlegendentry{TheSy\,%
    {\smaller(w\hspace{-1pt}/\hspace{-1pt}o case split)}}

  \end{axis}
\end{tikzpicture}

%% file: tables/cvc4-results.tex
\renewcommand\arraystretch{1.1}
\newcommand\cc[1]{~~#1~~}
\newcommand\ctwo[1]{\multicolumn{2}{c|}{~~#1~}}
\begin{tabular}{l|r|rrrr}
           & \cc{Total} & \cc{~~Z3} & \cc{CVC4} & \cc{CVC4+ig} & \cc{\TheSy} \\ \hline
clam       & 136   & 25 & 20~   & 108~~~        & 102    \\
hipspec    & 42    & 6  & 7~    & 33~~~         & 29    \\
isaplanner & 87    & 35 & 34~   & 79~~~         & 47    \\
leon       & 46    & 9  & 9~    & 40~~~         & 9     \\ \hline
Total      & 311   & 75 & 70~   & 260~~~        & 187  
\end{tabular}
\hspace{2pt}

%% file: img/cvc4-benchmarks/cactus-plot.tex
\begin{tikzpicture}
  \pgfplotsset{cactus/.style={
  		width=5cm,
        height=4cm,
        title style={yshift=-1pt,anchor=base},
        ytick distance=10^1,
        x tick label style={font={\smaller}},
        y tick label style={font=\tiny},
        ymajorgrids=true,
           mark size=.25pt}}

  \begin{axis}[title={clam}, cactus, ymode=log]
    \addplot[smooth,mark=*,green!50!gray] plot 
      table[y=time] {img/cvc4-benchmarks/cactus-clam.dat.txt};
  \end{axis}

  \begin{axis}[title={hipspec}, at={(5cm,0)}, cactus, ymode=log]
    \addplot[smooth,mark=*,green!50!gray] plot 
      table[y=time] {img/cvc4-benchmarks/cactus-hipspec.dat.txt};
  \end{axis}

  \begin{axis}[title={leon}, at={(0,-3.5cm)}, cactus, ymode=log]
    \addplot[smooth,mark=*,green!50!gray] plot 
      table[y=time] {img/cvc4-benchmarks/cactus-leon.dat.txt};
  \end{axis}

  \begin{axis}[title={isaplanner}, at={(5cm,-3.5cm)}, cactus, ymode=log]
    \addplot[smooth,mark=*,green!50!gray] plot 
      table[y=time] {img/cvc4-benchmarks/cactus-isaplanner.dat.txt};
  \end{axis}

\end{tikzpicture}

%% file: 07-related.tex
\section{Related Work}
\label{related}

\begin{paragraph}{Equality Graphs}~
Originally brought into use for automated theorem proving~\cite{JACM2005:Detlefs},
e-graphs were popularized as a mechanism for implementing low-level compiler
optimizations~\cite{POPL2009:Tate}, under the name \emph{PEGs}.
These e-graphs can be used to represent a large
program space compactly by packing together equivalent programs.
In that sense they are similar to Version Space 
Algebras~\cite{ML2003:Programming}, but their prime objective is entirely
different.
While VSAs focus on efficient intersections, e-graphs are used to saturate a
space of expressions with all equality relations that can be inferred.
They have found use in optimizing expressions for more than just speed,
for example to increase numerical stability of floating-point programs
in Herbie~\cite{PLDI2015:Panchekha}.
There are two key differences in the way e-graphs are used in this work compared
to prior:
(i) equality laws are not hard-coded nor fixed, they are fertilized
as the system proves more lemmas automatically;
(ii) saturation cannot be guaranteed or even obtained in all cases, which we overcome by a bound on rewrite-rule application depth.
(The latter point is an indirect consequence of the former.)
\end{paragraph}

\begin{paragraph}{Automated theorem provers}~
Many systems rely on known theorems or are designed to support users in semi-automated proving.
Congruence closure is also a proven method for tautology checking in
automated theorem provers, such as Vampire~\cite{CAV2013:Kovacs},
and is used as a decision procedure for reasoning about equality in leading SMT solvers
Z3~\cite{TACAS2008:DeMoura} and CVC4~\cite{CAV2011:Barrett}.
There, it is limited mostly to first-order reasoning, but can essentially
be applied unchanged to higher-level scenarios such as ours.

Related to theory exploration, but using separate techniques,
are Zipperposition~\cite{FroCoS2017:Cruanes}, and the conjecture generation mechanism implemented as part of
the induction prover in CVC4~\cite{VMCAI2015:Reynolds}.
It should be noted, that these are directed toward a specific proof goal, as opposed to theory exploration, which is presumed to be an offline phase.
As such, the above two techniques incorporate generation of inductive hypotheses into the saturation proof search / SMT procedure, respectively.
\end{paragraph}

\begin{paragraph}{Theory exploration}~
IsaCoSy~\cite{JAR2010:Johansson} pioneered the use of synthesis techniques for bottom-up lemma discovery.
IsaCoSy combines equivalence reduction with coun\-ter\-ex\-ample-guided inductive synthesis 
(CEGIS~\cite{ASPLOS2006/Solar-Lezama}) for filtering candidate lemmas.
This requires a solver capable of generating counterexamples to equivalence.
Subsequent development was based on random generation of test values, as implemented in QuickSpec~%
\cite{JFP2017:Smallbone} for reasoning about Haskell programs, later combined with automated provers
for checking the generated conjectures~\cite{ICAD2013:Claessen,ITP2017:Johansson}.
We have mentioned the deficiencies of using concrete values (as opposed to symbolic ones) and random testing in \autoref{intro} and 
make an empirical comparison with Hipster, a descendent of IsaCoSy and QuickSpec, in \autoref{evaluation}.
\end{paragraph}

\begin{paragraph}{Inductive synthesis}~
In the area of SyGuS~\cite{DSSE2015:Alur}, tractable bottom-up enumeration is commonly achieved
by some form of equivalence reduction~\cite{VMCAI2019:Smith}.
When dealing with concrete input-output examples, observational equivalence~%
\cite{CAV2103:Albarghouthi,Notices2013:Udupa} is very effective.
The use of symbolic examples in synthesis has been suggested~\cite{CAV2017:Drachsler}, but
to the best of our knowledge, ours is the only setting where symbolic observational equivalence
has been applied.
Inductive synthesis, in combination with abduction~\cite{STTT2017:Dillig},
has also been used to infer specifications~\cite{POPL2016:Albarghouthi},
although not as an exploration method but as a supporting mechanism for verification. 
\end{paragraph}

%% file: 09-conclusions.tex
\section{Conclusion}
\label{conclusions}

We described a new method for theory exploration, which differentiates itself from existing work by basing the reasoning on a novel engine based on term rewriting.
The new approach differs from previous work, specifically those based on testing techniques, in that:
\vspace{-.25em}
\begin{enumerate}
    \item This lightweight reasoning is purely symbolic, supporting value abstraction and performs better then prior art.
    \item Functions are naturally treated as first-class objects, without specific support implementation.
    \item The only needed input is the code defining the functions involved, and no support code such as a specific theory solver or random value generators.
    \item \TheSy has a unique feedback loop between the prover and the synthesizer, allowing more conjectures to be found and proofs to succeed.
\end{enumerate}
\vspace{-.25em}

By creating a feedback loop between the four different phases, term generation, conjecture inference, conjecture screening and induction prover, this system manages to efficiently explore many theories.
This goes beyond similar feedback loops in existing tools, aiming to reduce false and duplicate conjectures.
As explained in \autoref{overview:conjecture}, this form is also present in \TheSy, but \TheSy utilizes this feedback in more phases of the computation.

Theory exploration carries practical significance to many automated reasoning tasks,
especially in formal methods, verification and optimization.
 Complex properties lead to an ever-growing number of definitions and associated lemmas, which constitute an integral part of proof construction.
These lemmas can be used for SMT solving, automated and interactive theorem proving, and as a basis for equivalence reduction in enumerative synthesis.
The term rewriting-based method that we presented in this paper is simple, highly flexible, and has already shown results surpassing existing exploration methods.
The generated lemmas allow even this simple method to prove conjectures that normally require sophisticated SMT extensions.
Our main conclusion is that deductive techniques and symbolic evaluation can greatly contribute to theory exploration, in addition to their existing applications in invariant and auxiliary conjecture inference.

\subsubsection*{Acknowledgements.}
\label{sec:conclusion}

This research was supported by the Israeli Science Foundation (ISF) Grants No.~243/19 and
2740/19 and by the United States-Israel Binational Science Foundation (BSF) Grant
No.~2018675.